\documentclass[aps,prb,twocolumn,groupedaddress]{revtex4}

\usepackage{graphicx}
\usepackage{dcolumn}
\usepackage{bm}

\begin{document}

\preprint{APS/123-QED}

\title{Doping evolution of the electronic structure in the single-layer cuprates Bi$_2$Sr$_{2-x}$La$_x$CuO$_{6+\delta}$: Comparison with other single-layer cuprates}

\author{M. Hashimoto}
\affiliation{Department of Physics, University of Tokyo, Hongo, Tokyo 113-0033, Japan}
\author{T. Yoshida}
\affiliation{Department of Physics, University of Tokyo, Hongo, Tokyo 113-0033, Japan}
\author{H. Yagi}
\affiliation{Department of Physics, University of Tokyo, Hongo, Tokyo 113-0033, Japan}
\author{M. Takizawa}
\affiliation{Department of Physics, University of Tokyo, Hongo, Tokyo 113-0033, Japan}
\author{A. Fujimori}
\affiliation{Department of Physics, University of Tokyo, Hongo, Tokyo 113-0033, Japan}
\author{M. Kubota}
\affiliation{Photon Factory, Institute of Materials Structure Science, High Energy Accelerator Research Organization (KEK), Tsukuba, Ibaraki, 305-0801, Japan}
\author{K. Ono}
\affiliation{Photon Factory, Institute of Materials Structure Science, High Energy Accelerator Research Organization (KEK), Tsukuba, Ibaraki, 305-0801, Japan}
\author{K. Tanaka}
\affiliation{Department of Physics, Applied Physics, and Stanford Synchrotron Radiation Laboratory, Stanford University, Stanford, California 94305, U.S.A.}
\author{D.H. Lu}
\affiliation{Department of Physics, Applied Physics, and Stanford Synchrotron Radiation Laboratory, Stanford University, Stanford, California 94305, U.S.A.}
\author{Z.-X. Shen}
\affiliation{Department of Physics, Applied Physics, and Stanford Synchrotron Radiation Laboratory, Stanford University, Stanford, California 94305, U.S.A.}
\author{S. Ono}
\affiliation{Central Research Institute of Electric Power Industry, Komae, Tokyo 201-8511, Japan.}
\author{Yoichi Ando}
\affiliation{Institute of Scientific and Industrial Research, Osaka University, Ibaraki, Osaka 567-0047, Japan}

\date{\today}

\begin{abstract}
We have performed angle-resolved photoemission and core-level x-ray photoemission studies of the single-layer cuprate Bi$_2$Sr$_{2-x}$La$_x$CuO$_{6+\delta}$ (Bi2201) and revealed the doping evolution of the electronic structure from the lightly-doped to optimally-doped regions. 
We have observed the formation of the dispersive quasi-particle band, evolution of the Fermi ``arc'' into the Fermi surface and the shift of the chemical potential with hole doping as in other cuprates.
The doping evolution in Bi2201 is similar to that in Ca$_{2-x}$Na$_{x}$CuO$_{2}$Cl$_2$ (Na-CCOC), where a rapid chemical potential shift toward the lower Hubbard band of the parent insulator has been observed, but is quite different from that in La$_{2-x}$Sr$_{x}$CuO$_{4}$ (LSCO), where the chemical potential does not shift, yet the dispersive band and the Fermi arc/surface are formed around the Fermi level already in the lightly-doped region.
The (underlying) Fermi surface shape and band dispersions are quantitatively analyzed using tight-binding fit, and the deduced next-nearest-neighbor hopping integral $t'$ also confirm the similarity to Na-CCOC and the difference from LSCO.
\end{abstract}


\maketitle
\section{Introduction}

How the electronic structure of the antiferromagnetic insulator evolves into that of the superconductor with hole doping in the high-$T_c$ cuprates has been a major and fundamental issue in condensed-matter physics.
In the doping range where the insulator-to-superconductor transition occurs, dramatic changes occur in the thermodynamic and transport properties \cite{AndoCarrier00, OnoEvolution03, AndoElectronic04, EisakiEffect04, FujitaEffect05}, and exotic phenomena such as the pseudogap \cite{LoeserExcitation96,DingNature96}, Fermi ``arc'' \cite{ShenNodal05, YoshidaSystematic06, YoshidaMetallic03, KanigelEvolution06}, stripe order \cite{TranquadaEvidence95, YamadaDoping98} and 4a$\times$4a order \cite{Hanaguri'checkerboard'04, ShenNodal05} have been reported.

So far systematic angle-resolved photoemission (ARPES) studies on the doping evolution from the lightly-doped to underdoped regions have been performed only for the two single-layer cuprate families La$_{2-x}$Sr$_{x}$CuO$_{4}$ (LSCO) \cite{InoDoping-dependent02, YoshidaMetallic03, YoshidaLow-energy07,DamascelliAngle-resolved03, ShenFully04} and Ca$_{2-x}$Na$_{x}$CuO$_{2}$Cl$_2$ (Na-CCOC) \cite{ShenNodal05, ShenFully04,DamascelliAngle-resolved03, RonningEvolution03,ShenMissing04,KohsakaAngle-Resolved03}. 
These studies have revealed several common features such as the evolution of the pseudogap in the antinodal region and the Fermi ``arc'' in the nodal region.
From these studies combined with chemical potential shift measurements using core-level photoemission spectroscopy \cite{InoChemical97, YagiChemical06}, two different kinds of doping evolution have emerged. 
In LSCO, upon hole doping, the quasiparticle (QP) peak immediately appears around the Fermi energy ($E_F$) while the chemical potential $\mu$ (namely, the $E_F$ position) is pinned in the underdoped region.
The formation of such metallic dispersion with slight doping in LSCO has been demonstrated by Sahrakorpi $et$ $al.$ \cite{SahrakorpiAppearance07}.
The lower Hubbard band (LHB) stays away from $\mu$. 
In Na-CCOC, the chemical potential is shifted toward the LHB upon hole doping and further doping continues to lower the chemical potential into the LHB, creating the QP band and the Fermi arc/surface.
The question of why LSCO and Na-CCOC exhibit such contrasting behaviors has not been understood.
It has been suggested that the next-nearest-neighbor hopping integral $t$' plays an important role in the different band dispersions, Fermi surface shapes and chemical potential shifts of LSCO and the double-layer cuprates Bi$_2$Sr$_{2}$CuCu$_2$O$_{8+\delta}$ (Bi2212) \cite{TanakaEffects04, TohyamaDoping03, ChouLow-energy06, YangLow-energy06}.
In fact, different $t$' values are fundamentally important to understand the material dependences of the cuprates as theoretically suggested \cite{TohyamaDoping03, YangLow-energy06, PavariniBand-Structure01}.
In order to elucidate the origin of the differences between LSCO and Na-CCOC and those between LSCO and Bi2212, studies of other single-layer cuprates that cover a wide doping range are necessary. 
In this study, therefore, we have performed ARPES and core-level x-ray photoemission (XPS) studies on another single-layer cuprate system Bi$_2$Sr$_{2-x}$La$_x$CuO$_{6+\delta}$ (La-doped Bi2201) and compared the result with those of LSCO and Na-CCOC.

In La-doped Bi2201, as the La concentration $x$ increases, the hole concentration $p$ decreases, as determined by the room-temperature Hall coefficient \cite{AndoCarrier00}. 
For $p$ \textgreater 0.04, Bi2201 shows metallic in-plane resistivity at room temperature, similar to other high-$T_c$ cuprates.
For $p$ \textless 0.04, Bi2201 shows insulating behavior even at room temperature \cite{OnoEvolution03}, stronger tendency toward charge localization than LSCO and YBa$_2$Cu$_3$O$_{7-\delta}$ (YBCO) \cite{AndoElectronic04}. 
The electrical resistivity of lightly doped Bi2201 (\textit{p} \textless 0.10) shows divergence at low temperatures despite the metallic resistivity at room temperature.
Since the hole concentration of Bi2201 can be changed from $p$ = 0.03 to 0.18, one can study the doping dependence of the electronic structure from the lightly-doped to slightly overdoped regions in order to address the above important issues, although the undoped $p$ = 0 sample is difficult to synthesize.
Therefore, the measurements of chemical potential shift and ARPES in Bi2201 and subsequent comparison of the results with those of LSCO and Na-CCOC are expected to uncover the universal versus material-dependent properties in the doping evolution of the electronic structure.

\section{Experiment}

\begin{table}
\begin{center}
\caption{Chemical compositions $x$, hole concentration $p$ and $T_c$ of Bi$_2$Sr$_{2-x}$La$_x$CuO$_{6+\delta}$ samples studied in the present work.}
\label{Bi2201samples}
\begin{tabular}{lll}
\hline
\hline
$x$~~~~~~~~~~ & $p$~~~~~~~~~~ & $T_c$ (K)\\
\hline
0.96 & 0.05 & - \\
0.92 & 0.07 & - \\
0.91 & 0.07 & -  \\
0.80 & 0.10 & - \\
0.63 & 0.12 & 14\\
0.50 & 0.14 & 24 \\
0.40 & 0.16 & 34 \\
0.20 & 0.18 & 25 \\
\hline
\hline
\end{tabular}
\end{center}
\end{table}

High quality single crystals of La-Bi2201 were grown by the floating zone (FZ) method. 
Details of the sample preparation are described elsewhere \cite{OnoEvolution03, AndoCarrier00}. 
The La concentrations $x$, $p$ and $T_c$ of the measured samples are listed in Table \ref{Bi2201samples}.
The $p$ $\sim$ 0.10 sample was on the border of the insulator-to-superconductor transition.

ARPES measurements were performed at beamline 5-4 of Stanford Synchrotron Radiation Laboratory (SSRL), using a SCIENTA SES-200 analyzer with the total energy resolution of $\sim$15 meV and the angular resolution of 0.3 degree. 
Measurements were performed with the photon energy $h\nu$ = 19 eV and the polarization angle $(\pi/4)$ to the Cu-O bond.
The sample temperature was $\sim$10 K, which was below $T_c$ of the $p$ $\sim$ 0.12 sample.
The samples were cleaved \textit{in situ} under an ultrahigh vacuum of 10$^{-11}$ Torr. 
The Fermi edge of gold was used to determine the $E_F$ position and the instrumental resolution. 
ARPES measurements were also performed at beamline 28A of High Energy Accelerator Research Organization-Photon Factory (KEK-PF), using a SCIENTA SES-2002 analyzer with the total energy resolution of $\sim$25 meV and the angular resolution of 0.3 degree. 
Measurements were performed with the photon energy $h\nu$ = 55 eV and the circular polarization.
The sample temperature was $\sim$ 15 K.
Samples were cleaved \textit{in situ} under an ultrahigh vacuum of 10$^{-10}$ Torr. 

XPS measurements of core levels were performed using a SCIENTA SES-100 analyzer and an x-ray source of the Mg $K\alpha$ line (1253.6 eV) with the total energy resolution of $\sim$ 800 meV. 
The detailed experimental setup is described elsewhere \cite{YagiChemical06}.
Measurements were performed at 100 K. Samples were cleaved \textit{in situ} in 10$^{-10}$ Torr to obtain clean surfaces. 

\section{Results and discussion}
\subsection{Doping evolution of (underlying) Fermi surface}

\begin{figure}
\begin{center}
\includegraphics[width=9cm]{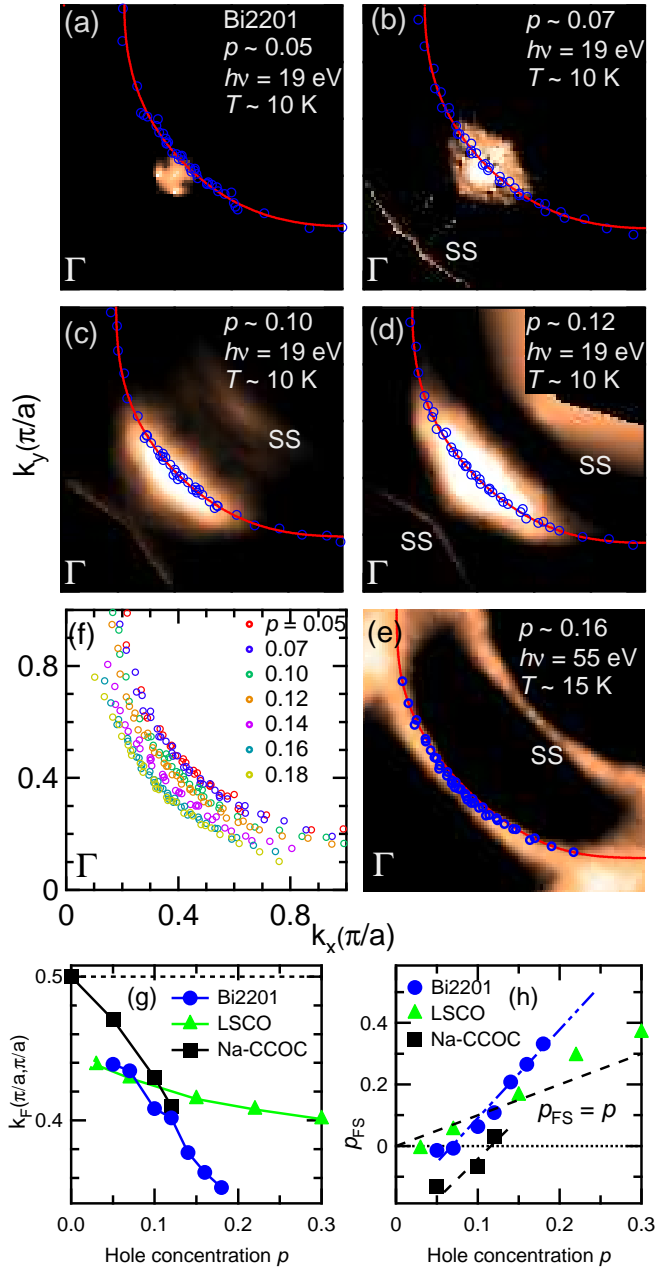}
\caption{(Color online) Doping dependence of the Fermi surface and ``underlying'' Fermi surface in Bi2201.
(a)-(e) $k$-space mapping of spectral weight at $E_F$ $\pm $30 meV window from lightly-doped to optimally-doped Bi2201 measured in the second Brillouin zone (BZ).
Super structures due to the Bi-O modulation are noted in the panels as SS.
Blue open circles indicate the $k_F$ positions determined by the peak positions of MDC's, both from the first and second BZ's.
Red curves show the results of tight-binding fit described below.
(f) Doping dependence of the $k_F$ positions, i.e., of the (underlying) Fermi surface.
(g) Doping dependence of the Fermi momentum $k_F$ in the nodal (0, 0)-($\pi$, $\pi$) direction.
(h) Apparent doping level defined by $p_{FS}$ = 2$S_{FS}/S_{BZ}-1$.
$p_{FS}$ = $p$ if Luttinger's sum rule is fulfilled.
Data for Na-CCOC (Ref. \onlinecite{ShenNodal05}) and LSCO (Ref. \onlinecite{YoshidaSystematic06}) are also plotted.
}
\label{FS_sym}
\end{center}
\end{figure}

In Fig.~\ref{FS_sym}(a)-(e), spectral weight within the $E_F$ $\pm$ 30 meV window is plotted in $k$-space for each doping level, showing the evolution of the Fermi arc/surface with doping.
All the intensities have been normalized to the second order light of the monochromator above $E_F$. 
A set of Fermi momenta $k_F$, which define the Fermi surface and the ``underlying'' Fermi surface, has been determined from the peak positions of the momentum distribution curves (MDC's) at $E_F$, and is plotted by open circles in the same panels.
Where a gap or a pseudogap is opened, $k_F$ has been determined by extrapolating the MDC peak to $E_F$.
As shown in Fig.~\ref{FS_sym}, the Fermi ``arc'' grows with doping.
For the insulating $p$ = 0.05, although a gap is opened on the entire Fermi surface, one can see a tiny intensity around the node due to the finite integration window.
For $p$ = 0.07, which is still insulating, the intensity around the node becomes higher than that for $p$ = 0.05.
For further hole doping, the $E_F$ intensity around the node increases, the actual Fermi ``arc'', where the QP peak crosses $E_F$, appears around the node at $p$ = 0.10.
The arc length becomes longer in going to the superconducting $p$ = 0.12 sample.
Although the overall picture of the doping evolution of the Fermi arc is similar to those of both LSCO \cite{InoDoping-dependent02, YoshidaLow-energy07, YoshidaSystematic06} and Na-CCOC \cite{ShenNodal05}, the details are more similar to that of Na-CCOC than to that of LSCO.
In LSCO, a distinct Fermi arc with the $E_F$ crossing of the QP peak appears already in the lightly-doped non-superconducting $x$ = 0.03 sample \cite{YoshidaMetallic03}, and this metallic feature with slight hole doping has been simulated theoretically \cite{SahrakorpiAppearance07}.
On the other hand, in the non-superconducting samples of Bi2201 ($p$ = 0.05, 0.07), the Fermi arc cannot be well defined because of the nodal gap opening as in Na-CCOC ($p$ = 0.05) \cite{ShenFully04}.

The $k_F$ positions for all doping levels are overlaid in Fig.~\ref{FS_sym}(f).
One can see that the hole-like Fermi surface (or underlying Fermi surface) is uniformly expanded with hole doping, quite different behavior from LSCO [Ref.~\onlinecite{YoshidaSystematic06}, reproduced in Fig.~\ref{ES}(d)].
The doping dependence of the $k_F$ position in the nodal direction is plotted in Fig.~\ref{FS_sym}(g) together with those for LSCO \cite{YoshidaSystematic06} and Na-CCOC \cite{ShenNodal05}.
While $k_F$ becomes smaller (closer to the $\Gamma$ point) with doping in every system, $k_F$ for Na-CCOC shows the strongest $p$ dependence and approaches ($\pi$/2, $\pi$/2) with underdoping \cite{ShenNodal05} while that for LSCO shows the weakest $p$ dependence and does not approach ($\pi$/2, $\pi$/2) with underdoping \cite{YoshidaSystematic06}.
The $p$ dependence of $k_{F}$ for Bi2201 is strong as in the case of Na-CCOC, but different from LSCO.
As we shall see below, in Bi2201 and Na-CCOC \cite{YagiChemical06}, the chemical potential is shifted upon hole doping, approaches the top of the LHB, while in LSCO \cite{InoChemical97, YoshidaSystematic06}, it stays away from the LHB in the underdoped region.
Thus, $k_{F}$ for Bi2201 and Na-CCOC \cite{ShenNodal05} show continuous shifts with doping from that of the antiferromagnetic parent compound, $\sim (\pi/2, \pi/2)$, while in LSCO \cite{YoshidaSystematic06}, for a slight amount of hole doping ($x$ = 0.03), $k_{F}$ is already at $\sim$(0.44$\pi$, 0.44$\pi$), away from antiferromagnetic BZ, and then shows only a weak doping dependence.  
This suggests that, in Bi2201 like Na-CCOC, the Fermi arc/surface evolve continuously from the top of the LHB, unlike LSCO, where the Fermi arc/surface is formed away from the LHB. 

We also examine the doping dependence of the (underlying) Fermi surface volume and compared it with those of LSCO \cite{YoshidaSystematic06} and Na-CCOC \cite{ShenNodal05} as shown in Fig.~\ref{FS_sym}(h).
Here, the precise Fermi surface volume $S_{FS}$ has been estimated using the tight-binding (TB) fit to the $k_F$ points, and the apparent doping level $p_{FS}$ is deduced by $p_{FS}$= 2$S_{FS}$/$S_{BZ}$-1, where $S_{BZ}$ = $4\pi ^2/a^2$ is the area of the Brillouin zone.
In this definition, $p_{FS}$ = $p$ means that Luttinger's sum rule is fulfilled \cite{SensarmaCan07}.
In LSCO, Luttinger's sum rule is approximately satisfied as shown in Fig. \ref{FS_sym}(h) (Ref.\onlinecite{YoshidaSystematic06}), however, the figure shows that $p_{FS}$ for Bi2201 significantly deviates from $p_{FS}$ = $p$ unlike LSCO \cite{YoshidaSystematic06} but similar to Na-CCOC \cite{ShenNodal05}.
In Bi2201 and Na-CCOC \cite{ShenNodal05}, therefore, there is a clear deviation from Luttinger's sum rule if the present definition of the (underlying) Fermi surface is employed.
Because in the underdoped region, there is a pseudogap, namely, there is no full Fermi surface, it is not obvious whether Luttinger's sum rule should be fulfilled or not.
The deviation of $p_{FS}$ from $p$ has been suggested by numerical simulations \cite{GrosDetermining06, SensarmaCan07}, which may explain the present results.
Here, the simulations have been done within renormalized mean field theory and it has been suggested that the sign of the deviation from $p$ is related to the Fermi surface topology.

\subsection{Doping evolution of chemical potential}

\begin{figure}
\begin{center}
\includegraphics[width=9cm]{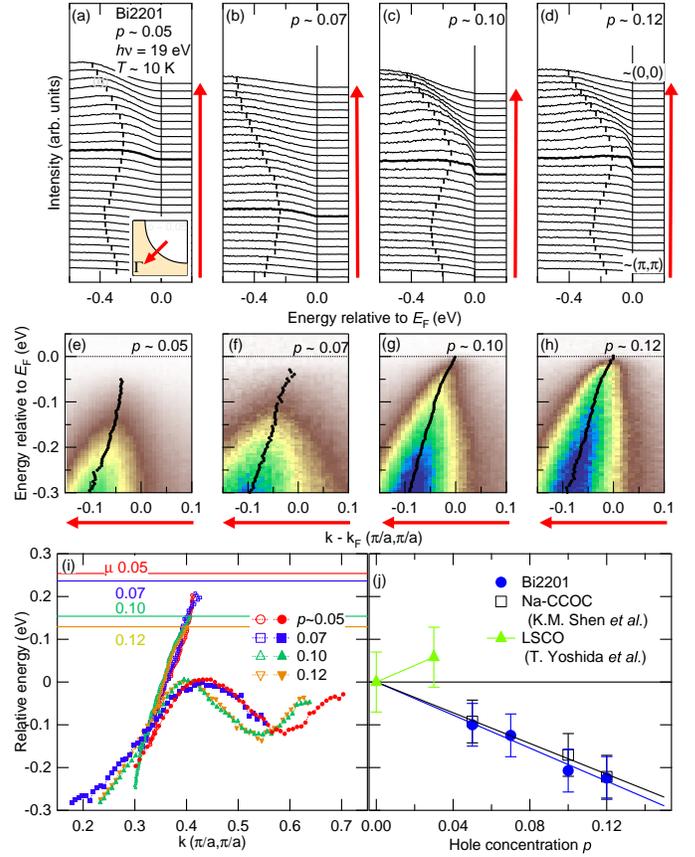}
\caption{(Color online) Doping dependence of the ARPES spectra along the nodal (0, 0)-($\pi$, $\pi$) direction for Bi2201 in the second BZ.
(a)-(d) Energy distribution curves (EDC's).
Thick curves show the EDC's at $k_F$.
Peaks in the second derivatives of the EDC's are marked and represent the lower Hubbard band (LHB). 
(e)-(h) Intensity plots in energy-momentum ($E$-$k$) space.
Black curves show the MDC peak positions and define the quasi-particle (QP) band dispersion.
(i) Relative positions of the LHB and the QP band.
They have been shifted vertically so that the top of the LHB's for the different doping levels are aligned.
The horizontal lines show the chemical potential $\mu$ for the various doping levels.
(j) Shift of the chemical potential relative to the LHB of Bi2201 and that of Na-CCOC (Ref. \onlinecite{ShenMissing04}) and LSCO (Ref. \onlinecite{YoshidaMetallic03}).
The plots have been shifted so that the extrapolated value to zero doping is aligned.
}
\label{Node}
\end{center}
\end{figure}

The doping evolution of the ARPES spectra along the nodal (0, 0)-($\pi$, $\pi$) direction from the lightly-doped to underdoped Bi2201 are shown in Fig.~\ref{Node}.
Panels (a)-(d) show energy distribution curves (EDC's) and panels (e)-(h) show energy-momentum ($E$-$k$) intensity plots.
The peaks in MDC's marked by black curves represent the ``QP band'' dispersion.
For the most lightly-doped $p$ = 0.05, the intensity around $E_F$ is very weak and there is no QP band crossing the chemical potential.
The LHB position determined from the second derivative of the EDC's, as in the previous work on Na-CCOC \cite{ShenMissing04}, is marked in panels (a)-(d).
The top of the LHB is located at $\sim$ -0.25 eV in the $p$ = 0.05 sample.
As the doping level increases, the LHB approaches $E_F$ and the intensity around $E_F$ gradually increases.
For $p$ = 0.10 and 0.12, the QP reaches $E_F$ and crosses it as shown in Fig \ref{Node}(g) and (h).

In order to clarify the relationship between the shifts of the LHB, the QP band and the chemical potential $\mu$ with hole doping, we plot in Fig.~\ref{Node}(i) $\mu$ and the QP dispersion referenced to the LHB.
That is, the positions of the LHB and the QP band have been shifted vertically so that the top of the LHB is aligned.
One can clearly see that the chemical potential gradually moves downward with doping.
Interestingly, the QP band did not move with doping in this plot, meaning that the QP band structure exhibits a rigid-band behavior as in the case of Na-CCOC \cite{ShenMissing04}.
Also, the Fermi velocity is almost constant, $v_F$ = 1.8 eVA$^{-1}$, similar to the previous studies on LSCO \cite{YoshidaSystematic06} and Na-CCOC \cite{ShenMissing04}, reflecting the ``universal nodal Fermi velocity'' \cite{ZhouUniversal03}.
The shift of $\mu$ relative to the LHB in Bi2201 is as fast as that in Na-CCOC \cite{ShenMissing04} and is much faster than that in LSCO \cite{YoshidaMetallic03} as plotted in Fig.~\ref{Node}(j).

\begin{figure}
\begin{center}
\includegraphics[width=9cm]{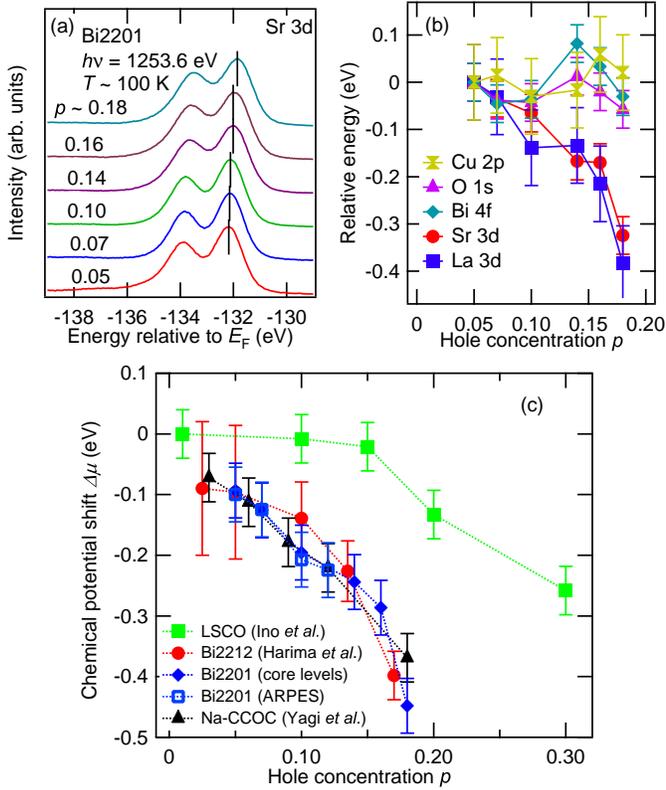}
\caption{(Color online) Core-level photoemission results for Bi2201.
(a) Photoemission spectra of the Sr 3$d$ core level from $p$=0.05 to 0.18 for Bi2201.
(b) Doping dependence of each core level peak relative to $p$ = 0.05.
(c) Chemical potential shift $\Delta\mu$ in Bi2201 compared with those in LSCO (Ref.\onlinecite{InoChemical97}), Bi2212 (Ref.\onlinecite{HarimaChemical03}) and Na-CCOC (Ref.\onlinecite{YagiChemical06}).}
\label{coreshift}
\end{center}
\end{figure}

In order to deduce the chemical potential shift with respect to a more stable reference than the LHB, which could change its position and dispersion with hole doping, we have measured core-level XPS spectra of Bi2201 as a function of doping as shown in Fig.~\ref{coreshift}.
Panel (a) shows the doping dependence of the Sr 3$d$ core levels and panel (b) shows the shifts of various core levels as a function of hole doping.
As can be seen from panel (b), the Sr 3$d$ and La 3$d$ core levels are shifted monotonously with doping, while the other core levels show complicated behaviors.
Since these results are very similar to those of Bi2212 \cite{HarimaChemical03}, we have performed the same analysis for Bi2201 as done for Bi2212, assuming that the shifts of the Sr 3$d$ and La 3$d$ core levels reflect the chemical potential shift $\Delta\mu$.
In Fig.~\ref{coreshift}(c), we have plotted thus deduced $\Delta\mu$ of Bi2201 as a function of hole concentration together with those of other cuprates.
The chemical potential of Bi2201 shows a monotonous shift of $\partial\mu / \partial p$ $\sim$ -1.7 eV/hole.
One can clearly see that the shift of Bi2201 is similar to those of Na-CCOC \cite{YagiChemical06} and  Bi2212 \cite{HarimaChemical03}, but is very different from that of LSCO \cite{InoChemical97}, which shows pinning behavior $\partial\mu / \partial p$ $\sim$ 0 eV/hole in the underdoped region.
The results indicate a similarity between Bi2201 and Na-CCOC and dissimilarity from LSCO.
The calculation using the $t-t'-t''-J$ model has indicated that the shift becomes faster with increasing $-t'$ \cite{TohyamaDoping03}.
The observed shifts, therefore, suggest that the $-t'$ of Bi2201 is similar to those of Na-CCOC and Bi2212, and is larger than that of LSCO.

\subsection{Doping evolution of band dispersion}

\begin{figure}
\begin{center}
\includegraphics[width=9cm]{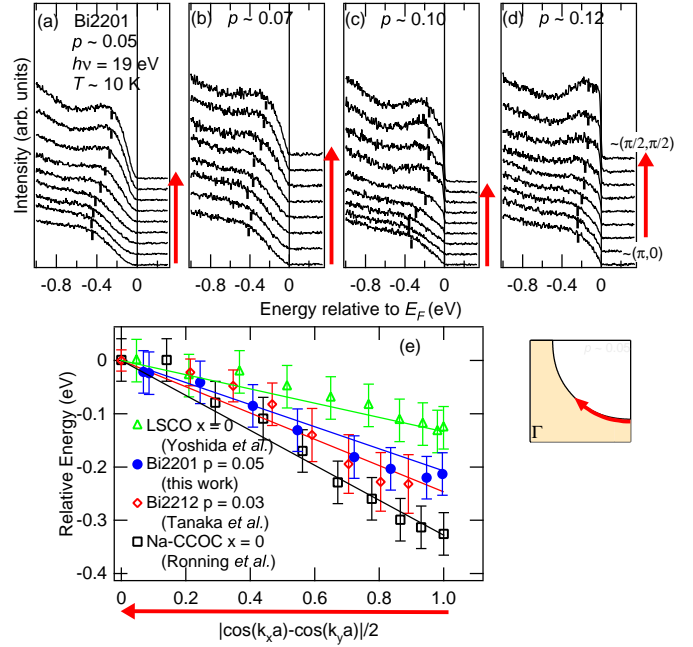}
\caption{(Color online) ARPES spectra of Bi2201 along the (underlying) Fermi surface.
(a)-(d) EDC's along the (underlying) Fermi surface in the second BZ of Bi2201.
Vertical bars show the LHB positions determined by the second derivatives of the EDC's.
(e) Shift of the LHB against the $d$-wave function (\textbar$ \cos(k_xa)$-$cos(k_ya)$\textbar/2) for the lowest doping $p$ = 0.05 compared with those of La$_2$CuO$_4$ (Ref. \onlinecite{TanakaEffects04}), CCOC (Ref. \onlinecite{RonningPhotoemission98}) and Bi2212 (Ref. \onlinecite{TanakaEffects04}).
}
\label{arc}
\end{center}
\end{figure}

The EDC's shown in Fig.~\ref{arc}(a)-(d) show band dispersion along the (underlying) Fermi surface in Bi2201.
For all the doping levels, the dispersive feature is marked by vertical bars determined by the second derivatives of the EDC's, and becomes deepest at $\sim$($\pi$, 0) and closest to $\mu$ at $\sim$($\pi$/2, $\pi$/2).
This dispersion together with that shown in Fig. \ref{Node}(a)-(d) represent the dispersion of the LHB, indicating that the LHB shows the maximum around  $\sim$($\pi$/2, $\pi$/2).
With hole doping, the LHB moves upward and the QP peak appears around the node for $p$ \textgreater 0.10.
The position and the doping dependence of the LHB are similar to those in Na-CCOC \cite{ShenNodal05}, but are quite different from those in LSCO \cite{InoDoping-dependent02}, where the top of the LHB stays $\sim$0.5 eV below \textit{E$_F$} until it fades out at higher doping $\ge$ 0.10.
In the antinodal region, the dispersive feature moves from $\sim$-0.45 eV for $p$ = 0.05 to $\sim$-0.25 eV for $p$ = 0.12 but does not approach $\mu$ further, clearly indicating that the so-called ``large pseudogap'' is opened around the antinodal region in the underdoped samples as in the other single-layer cuprates, LSCO \cite{InoDoping-dependent02, YoshidaLow-energy07, ShenFully04} and Na-CCOC \cite{ShenNodal05}.
From the dispersional width of the LHB from $\sim(\pi/2, \pi/2)$ to $\sim(\pi, 0)$ in the low doping limit, the magnitude of $t'$ can be estimated \cite{KimSystematics98}.
As shown in Fig. \ref{arc}(e), the dispersional width for Bi2201 is $\sim$0.20 eV, which is smaller than that for CCOC \cite{RonningPhotoemission98} ($\sim$0.35 eV) but comparable to that for Bi2212 \cite{TanakaEffects04}, and much larger than La$_2$CuO$_4$ (Ref. \onlinecite{TanakaEffects04}), consistent with the increase of $t'$ in going from LSCO to the other cuprates, estimated from the TB fit as discussed below.

\begin{figure}
\begin{center}
\includegraphics[width=9cm]{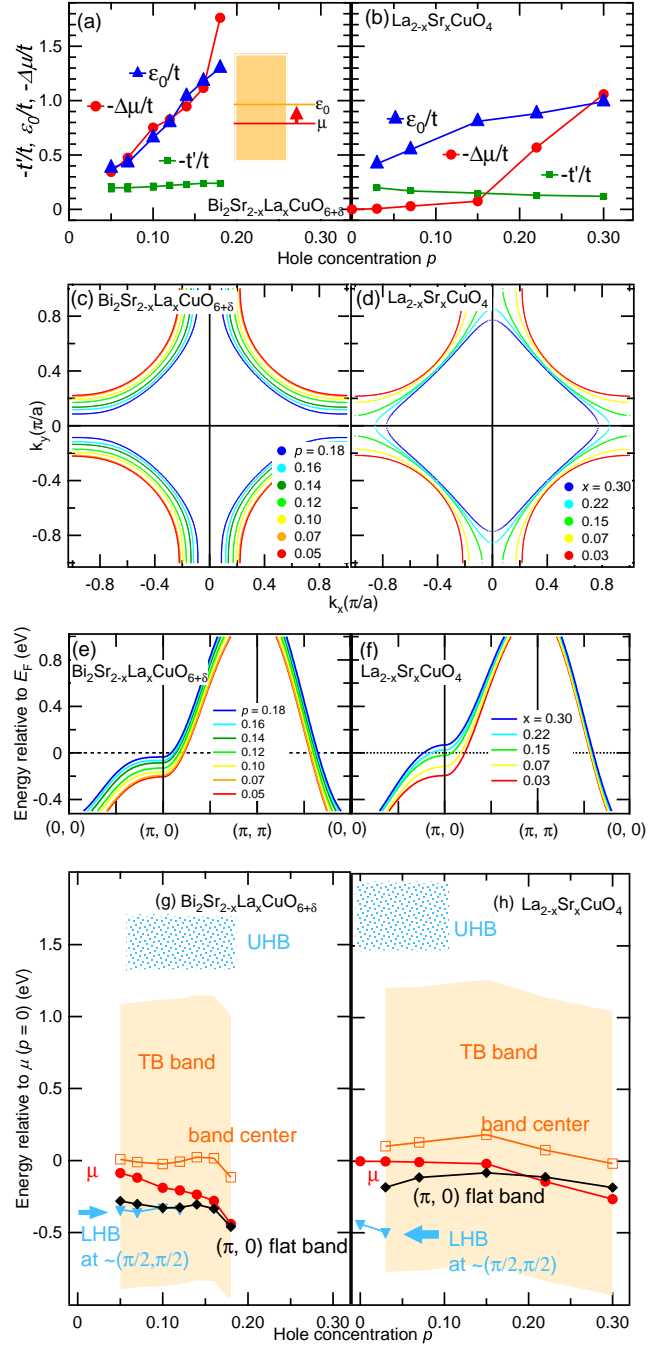}
\caption{(Color online) Doping dependence of the electronic structure in Bi2201 and LSCO.
(a)(b) Doping dependence of the tight-binding (TB) parameters and -$\Delta \mu/t$ determined from the core-level shifts.
The parameters for LSCO are taken from Refs.\onlinecite{YoshidaLow-energy07} and \onlinecite{InoChemical97}.
The inset to panel (a) shows the definition of $\epsilon _0$.
(c)(d) Doping dependence of the (underlying) Fermi surface shape.
That for LSCO has been taken from Ref.\onlinecite{YoshidaLow-energy07}.
(e)(f) Band dispersion from the TB fit.
The TB parameters for LSCO are taken from Ref.~\onlinecite{YoshidaSystematic06}.
(g)(h) Entire picture of the doping dependence of the electronic structure.
LSCO are taken from (Ref.~\onlinecite{YoshidaSystematic06}).
The approximate position of the upper Hubbard band (UHB) has been taken from inverse-photoemission spectra (Refs.~\onlinecite{WagenerComparison90,WatanabeInverse-photoemission91,OhtaInverse89,FujimoriChemical98}).
The energy range of the TB band dispersion is shown by orange region.
}
\label{ES}
\end{center}
\end{figure}

In order to evaluate the doping dependence of the shape of the (underlying) Fermi surface and the QP band dispersion more quantitatively, we have fitted the ARPES results to the two-dimensional single-band TB model,
\begin{eqnarray*}
\epsilon_k - \mu = -2t[\cos(k_xa) + \cos(k_ya)] - 4t'\cos(k_xa)\cos(k_ya)\\
   - 2t''[\cos(2k_xa) + \cos(2k_ya)] + \epsilon_0.
\label{eq1}
\end{eqnarray*}
We have assumed the relationship $t''$/$t'$ = -1/2 as before \cite{YoshidaSystematic06, Lee0606347}, and regarded -$t'$/$t$ and -$\epsilon_0$/$t$ as fitting parameters.
Here, $\epsilon_0$ is the position of the band center relative to the chemical potential $\mu$ [see the inset to Fig.~\ref{ES}(a)].
If the chemical potential shift $\Delta \mu$ is entirely due to rigid-band like one, $\Delta \mu/t$ = -$\epsilon_0$/$t$.
The value of $-t'/t$ can be estimated from the shape of the Fermi surface.
If $-t'/t$ is large, the Fermi surface becomes ``square-like'' and if $-t'/t$ is small, the Fermi surface becomes ``diamond-like'' \cite{PrelovifmmodeSpectral02} [see Fig. \ref{ES}(c) and (d)].
The shape of the (underlying) Fermi surface for Bi2201 is nearly circular and is therefore less ``diamond-like'' than that of LSCO \cite{YoshidaLow-energy07, YoshidaSystematic06}, similar to Na-CCOC \cite{ShenNodal05}, and therefore $-t'/t$ for Bi2201 should be larger than that for LSCO.
The fitted (underlying) Fermi surfaces drawn in Fig.~\ref{FS_sym}(a)-(e) by red curves are reproduced and overlaid in Fig. \ref{ES}(c) and (d).
The doping dependence of the fitted parameters are shown and compared with those of LSCO in Fig. \ref{ES}(a) and (b).
Here, we used $t$ = 0.25 eV determined from the velocity in the nodal direction.
While -$\epsilon_0$/$t$ increases with doping corresponding to the hole doping, -$t'$/$t$ shows only weak doping dependence as shown in Fig.\ref{ES}(a).
The larger slope of $-\epsilon _0/t$ in Bi2201 than that in LSCO \cite{YoshidaSystematic06} means that $\mu$ is shifted faster relative to the band center in Bi2201 than in LSCO.
This corresponds to the lower density of states of QP's at $\mu$ in Bi2201 than in LSCO, that is, the larger Fermi velocity near $\sim$($\pi$, 0) in Bi2201 than in LSCO, as can be seen from the band dispersion in Fig~\ref{ES}(e) and (f).

According to LDA calculations, the larger the Cu-to-apical oxygen distances is, the larger -$t'/t$ is \cite{PavariniBand-Structure01}. 
The present results are quantitatively consistent with the LDA calculations, where -$t'$/$t$ for Bi2201 is larger than that for LSCO and smaller than those for Hg and Tl based cuprates.
The LDA calculation \cite{PavariniBand-Structure01} has also indicated that -$t'/t$ strongly depends on the Cu-apical oxygen distance in LSCO, but not in Bi2201.
This is consistent with the present experimental result that $t'/t$ for Bi2201 shows much weaker doping dependence than that for LSCO \cite{YoshidaSystematic06}.
We note that the suggestion has been made that -$t'/t$  is correlated with the observed maximum $T_{c,max}$ \cite{TanakaEffects04, PavariniBand-Structure01}.
From the present work, however, the correlation between -$t'/t$ and $T_{c, max}$ is not clear since $T_{c, max}$ is similar for Bi2201, LSCO and Na-CCOC in spite of the different $-t'/t$.
One possibility is the strong disorder effects from out of the CuO$_2$ planes, which is strong in single-layer cuprates \cite{EisakiEffect04, FujitaEffect05}, but is not reflected on $-t'/t$.
This has to be clarified in future studies.
 
The doping evolutions of the band dispersion determined from the TB fit are shown in Fig.~\ref{ES}(e) and (f).
The result for Bi2201 demonstrates a rigid-band like evolution, i.e., a uniform shift of the entire band dispersion, while that for LSCO \cite{YoshidaSystematic06} shows different behavior, i.e., larger shift around $(\pi, 0)$ than along the $(0, 0)$-$(\pi, \pi)$ line.
The similarity between $\epsilon_0$/$t$ and -$\Delta \mu/t$ in Bi2201 [Fig.~\ref{ES}(a)], also confirm the rigid-band-like behavior.
The deviation from the rigid-band behavior in LSCO is attributed to the strong doping dependence of $t'$, which affect the antinodal region significantly.
As a result, the topology change of the Fermi surface from the hole-like to electron-like with hole doping is accelerated in LSCO [Fig. \ref{ES}(c) and (d)].

In Fig.~\ref{ES}(g) and (h), the chemical potential $\mu$, the band center ($\epsilon_0+\mu$), the ($\pi$, 0) flat band position, the top of the LHB [at $\sim (\pi/2, \pi/2)$] and the upper Hubbard band (UHB) \cite{WagenerComparison90,WatanabeInverse-photoemission91,OhtaInverse89, FujimoriChemical98} positions are plotted relative to the chemical potential at $p$ = 0.
The range of the TB band is also shown by the shaded area.
If one concentrates on the doping dependence of $\mu$ and the band center, for example, one can see from Fig~\ref{ES} (g) and (h) several clear differences between Bi2201 and LSCO. For Bi2201, all quantities except for $\mu$ are relatively unchanged with doping until $p$ $\sim$ 0.16 and only $\mu$ moves downward with hole doping, that is, the typical rigid-band-like shift is realized.
On the other hand, in the underdoped LSCO, $\mu$ is pinned, while the other band energy positions change with doping. 
In the overdoped region, the evolution becomes rather close to rigid-band-like. 
Although there are differences how the dispersion appears with hole doping to the Mott insulator between Bi2201 and LSCO, the metallic dispersion appears with hole doping, and the band is filled with hole doping in Bi2201, similar to LSCO \cite{SahrakorpiAppearance07, YoshidaMetallic03, YoshidaSystematic06}.
The present results indicate that the similar mechanism can be applied to the doping evolution of the dispersion in Bi2201, while the spectral weight is strongly suppressed with underdoping in Bi2201 compared to that for LSCO.
The origin of this difference between Bi2201 and LSCO remains an open question.

\section{Conclusion}

We have investigated the doping dependence of the electronic structure of the single-layer Bi2201 by ARPES and chemical potential shift measurements and found that the doping evolution is different from LSCO but is similar to Na-CCOC.
That is, in Bi2201, the doping evolution can be understood as a rigid-band-like shift of the chemical potential into the LHB, in contrast to LSCO, where the chemical potential is pinned well above the LHB in the under doped region as the QP band and the Fermi arc/surface are formed around the chemical potential.
The similarities to Na-CCOC and the differences from LSCO in the (underlying) Fermi surface shape and the QP band dispersions are accounted by the similar and different values of $t'$, respectively.

We have thus revealed that there are two kinds of doping evolution of the electronic structure in the underdoped cuprates.
In the present paper, we have not discussed why there are two different evolutions on a microscopic level beyond the differences in the parameter -$t'/t$ values.
Disorder effects are also material dependent and significantly affect the doping evolution and the $T_c$. 
The material dependent doping evolution may also be related with the formation of stripes versus 4a$\times $4a orders.
Other material-dependent effects such as electron-phonon coupling strength \cite{MishchenkoElectron-Phonon04,RoschPolaronic05} remain to be clarified in future studies.
To critically investigate how the electronic structure is affected by these material dependent factors will resolve the existence of the different doping evolutions.
In the overdoped metallic region, doping evolution would be relatively material-independent, in the sense that the rigid-band-like behavior is observed in LSCO, too, although the band structure of LSCO becomes different, that is, the ($\pi$,0) flat band moves to above $\mu$ and the Fermi surface topology changes from hole-like to electron-like.
The measurements on more overdoped Bi2201 are needed to carefully examine the material dependence in the overdoped region.

\section*{ACKNOWLEDGEMENT}

This work was supported by a Grant-in-Aid for Scientific Research in Priority Area ``Invention of Anomalous Quantum Materials'' from MEXT, Japan and by the US-Japan Joint Research Program from JSPS.
SSRL is operated by the Department of Energy's Office of Basic Energy Science, Division of Chemical Sciences and Material Sciences.
The work at KEK-PF was done under the approval of Photon Factory Program Advisory Committee (Proposal No. 2006S2-001) at the Institute of Material Structure Science, KEK.
Y. A. was supported by Grants-in-aid in Scientific Research 16340112 and 19674002.

\end{document}